\documentclass[pdflatex,referee,sn-standardnature]{sn-jnl}% Basic Springer Nature Reference Style/Chemistry Reference Style

\usepackage{graphicx}
\jyear{2023}%

\unnumbered% uncomment this for unnumbered level heads
\usepackage{natbib}
\begin{document}

\title[Ultrafast Hot ILX]{Time-domain observation of interlayer exciton formation and thermalization in a MoSe$_2$/WSe$_2$ heterostructure}


\author*[1,6]{\fnm{Veronica R.} \sur{Policht}}\email{veronica.policht.ctr@nrl.navy.mil}
\author*[2]{\fnm{Henry} \sur{Mittenzwey}}\email{h.mittenzwey@tu-berlin.de}
\author[1]{\fnm{Oleg} \sur{Dogadov}}
\author[2]{\fnm{Manuel} \sur{Katzer}}
\author[1]{\fnm{Andrea} \sur{Villa}}
\author[3]{\fnm{Qiuyang} \sur{Li}}
\author[4]{\fnm{Benjamin} \sur{Kaiser}}
\author[1]{\fnm{Aaron M.} \sur{Ross}}
\author[1]{\fnm{Francesco} \sur{Scotognella}}
\author[3]{\fnm{Xiaoyang} \sur{Zhu}}
\author[2]{\fnm{Andreas} \sur{Knorr}}%\email{andreas.knorr@tu-berlin.de}
\author[2]{\fnm{Malte} \sur{Selig}}
\author[1,5]{\fnm{Giulio} \sur{Cerullo}}
\author*[1]{\fnm{Stefano} \sur{Dal Conte}}\email{stefano.dalconte@polimi.it}

\affil[1]{\orgdiv{Department of Physics}, \orgname{Politecnico di Milano}, \orgaddress{\street{Piazza Leonardo da Vinci 32}, \city{Milano}, \postcode{20133}, \country{Italy}}}
\affil[2]{\orgdiv{Institut f\"ur Theoretische Physik, Nichtlineare Optik und Quantenelektronik}, \orgname{Technische Universität Berlin}, \orgaddress{Hardenbergstra{\ss}e 36}, \postcode{10623}, \city{Berlin}, \country{Germany}}
\affil[3]{\orgdiv{Department of Chemistry}, \orgname{Columbia University}, \orgaddress{\street{3000 Broadway}, \city{New York}, \postcode{10027}, \state{NY}, \country{United States}}}
\affil[4]{\orgdiv{Zuse-Institut Berlin}, \orgaddress{Takustraße 7}, \postcode{14195}, \city{Berlin}, \country{Germany}}
\affil[5]{\orgname{CNR-IFN}, \orgaddress{\street{Piazza Leonardo da Vinci 32}, \city{Milano}, \postcode{20133}, \country{Italy}}}
\affil[6]{\orgname{Current Address: U.S. Naval Research Laboratory}, \orgaddress{\street{4555 Overlook Avenue SW}, \city{Washington}, \state{DC}, \postcode{20375}, \country{USA}}}

%%==================================%%

\abstract{Vertical heterostructures (HS) of transition metal dichalcogenides (TMDs) host interlayer excitons (ILX), with electrons and holes residing in different layers. With respect to their intralayer counterparts, ILX feature much longer lifetimes and diffusion lengths, paving the way to excitonic optoelectronic devices operating at room temperature. While the recombination dynamics of ILX has been intensively studied, the formation process and its underlying physical mechanisms are still largely unexplored. Here we use ultrafast transient absorption spectroscopy with a white-light probe, spanning both intralayer and interlayer exciton resonances, to simultaneously capture and time-resolve interlayer charge transfer and ILX formation dynamics in a MoSe$_2$/WSe$_2$ HS. We find that the ILX formation timescale is nearly an order of magnitude ($\sim$1 ps) longer than the interlayer charge transfer time ($\sim$100 fs). Microscopic calculations attribute the relative delay to an interplay between a phonon-assisted interlayer exciton cascade and subsequent cooling processes, and excitonic wave-function overlap. Our results provide an explanation to the efficient photocurrent generation observed in optoelectronic devices based on TMD HS, as the ILX have an opportunity to dissociate during their thermalization process.}

\keywords{2D Materials, Interlayer Exciton, Transition Metal Dichalcogenide, Heterostructures}

\maketitle
\section{Introduction} \label{sec1}

Monolayer (ML) transition metal dichalcogenides (TMDs) exhibit remarkable physical properties which make them an ideal platform to study exciton physics and to realize novel optoelectronic devices \cite{Wang2012b}. The dimensional reduction from bulk to 2D results in the formation of strongly bound excitons due to increased quantum confinement and reduced Coulomb screening, along with spin-valley locking due to the inversion symmetry breaking \cite{Qiu2013a,Schaibley2016a}. The ability to vertically stack multiple TMDs, forming van der Waals heterostructures (HS) without lattice matching constraints, has dramatically increased recent interest in these materials \cite{Geim2013}. Despite the fact that weak out-of-plane interactions largely preserve the electronic structures of each layer, stacked HS display novel properties and functionalities not present in constituent monolayers. HS with Type II band alignment, where the valence band maximum and the conduction band minimum are in different layers, can host interlayer excitons (ILX) (Fig. \ref{Fig1}a) which arise following interlayer charge transfer (ICT) \cite{Hong2014b,Policht2021,Purz2021} and consist of spatially separated Coulomb-bound electron-hole states with binding energies up to 100s meV \cite{Jiang2021}. The ILX in TMD-HS is commonly detected via its photoluminescence (PL) in the near-infared (NIR), below the energy of the optical gap of the two layers (Fig. \ref{Fig1}b) \cite{Rivera2015,Nagler2017,Nayak2017,Hanbicki2018,Tran2019c}. The ILX is characterized by a long recombination time (up to hundreds of nanoseconds \cite{Wang2019,Jauregui2019}) and reduced oscillator strength (two orders of magnitude lower than that of intralayer excitons) due to the small spatial overlap of the electron and hole wave functions and, in some instances, their momentum-indirect character \cite{Rivera2015,Miller2017c,Okada2018,Hanbicki2018,Kunstmann2018}. One of the most intriguing properties of TMD-HS is the lateral confinement of the ILX within moiré potentials, which are formed at small interlayer twist angles \cite{Jin2019c,Alexeev2019b,Tran2019c,Seyler2019,Huang2022}.

ILX dynamics have been explored in different TMD-HS by time-resolved techniques \cite{Jiang2021}. The ILX recombination dynamics measured by time-resolved PL exhibit multiple decay components which exceed the recombination lifetimes of intralayer excitons by several orders of magnitude \cite{Rivera2015,Nagler2017a}. The ILX relaxation dynamics have also been indirectly inferred from the photobleaching (PB) dynamics of intralayer excitons as measured by transient absorption (TA) optical spectroscopy \cite{Wang2021}. On the other hand, the formation dynamics of the ILX are very difficult to access directly due to several factors, including the weak oscillator strength \cite{Ross2017c,Barre2022} and the rapidity of the ICT process that leads to ILX formation, which typically ranges from tens to hundreds of femtoseconds \cite{Hong2014b,Liu2020f,Policht2021,Wang2021,Zimmermann2021c,Purz2021}. Time-resolved PL lacks the temporal resolution required to observe these processes. Attempts to resolve the transition from intralayer excitons to ILX have been done by measuring the formation of a novel $\emph{1s-2p}$ transition in the mid-IR range \cite{Merkl2019}. More recently, time- and angle-resolved photoemission spectroscopy (tr-ARPES) experiments have addressed ILX formation dynamics \cite{Karni2022,Schmitt2022} by probing the energy-momentum dispersion of photoexcited quasiparticles in real time and distinguishing excitons from single particle states according to their energy-momentum dispersion \cite{Perfetto2016,Rustagi2018,Christiansen2019,Dong2021,Man2021,Schmitt2022}. A recent tr-ARPES study \cite{Schmitt2022} was able to track the ILX formation process following a phonon-assisted interlayer electron transfer as mediated by intermediate scattering to the $\Sigma$ valleys. Despite the strengths of this technique, the limited energy resolution and the extremely low intensity of the photoemission signal above the Fermi level complicates the ability to disentangle ILX formation from ICT dynamics \cite{Lee2021}.

Here we use ultrafast optical TA spectroscopy to directly probe the transient optical response of the ILX in a MoSe$_2$/WSe$_2$ HS. These measurements are enabled by highly stable and broadband white light probe pulses spanning the visible to the NIR. The resulting high signal-to-noise measurements are capable of resolving the ILX formation dynamics through its weak TA signal (two orders of magnitude lower than the TA of the intralayer exciton) while simultaneously measuring intralayer exciton and interlayer hole transfer (IHT) dynamics. The ILX signal shows a delayed growth on a picosecond timescale, which is significantly longer than the experimentally measured 100-fs IHT process from MoSe$_2$ to WSe$_2$. We simulate the exciton dynamics by solving the microscopic Heisenberg equations of motion and find that the difference in formation timescales is due to phonon-assisted hole tunneling of photo-excited excitons which gives rise to hot ILX populations that quickly exchange energy and momentum with phonons. The relaxation down to the ILX ground state proceeds through multiple scattering processes involving higher energy interlayer $\emph{s}$ states. Our simulations demonstrate that these hot ILX populations contribute strongly to the WSe$_2$ PB signal but only weakly to the optically bright ILX transition, resulting in the relative delay of the ILX signal as the populations cool.
 
\section{Results}\label{sec2}

\subsection{Experimental Results}\label{subsec21}

\begin{figure}[h!]%
\centering
\includegraphics[width=8.8cm]{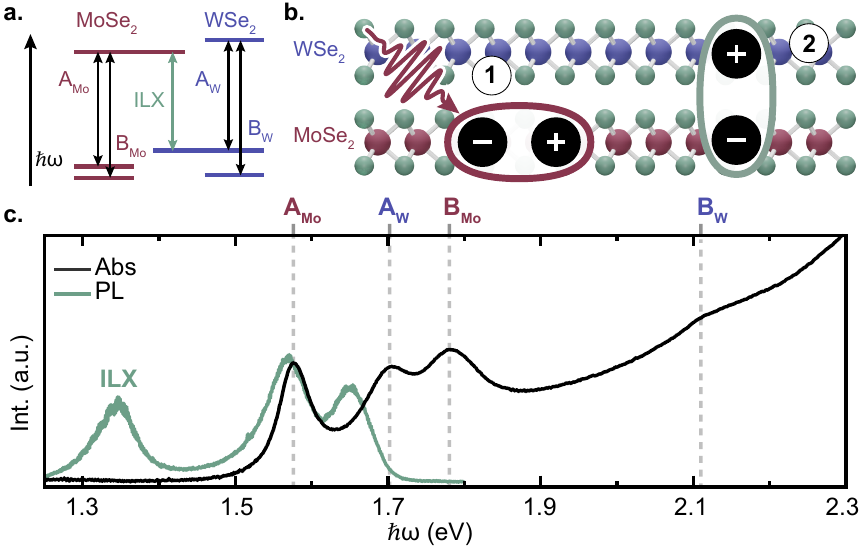}
\caption{\textbf{Interlayer Exciton formation in MoSe$_2$/WSe$_2$ HS.} \textbf{a} Type II band alignment diagram of the MoSe$_2$/WSe$_2$ HS. The relevant intralayer and interlayer exciton transitions are depicted with arrows. \textbf{b}, ILX formation in the MoSe$_2$/WSe$_2$ HS. An intralayer exciton is resonantly excited in the MoSe$_2$ layer (1). IHT to WSe$_2$ causes the formation an optically bright ILX (2). \textbf{c}, Linear absorption (black trace) and PL (green trace) spectra of the HS at 77 K. The four peaks in the linear absorption, marked by vertical dashed lines, correspond to A/B intralayer excitons. The PL spectrum reveals a spectral feature below the intralayer exciton absorption edge, which corresponds to ILX emission peak.}\label{Fig1}
\end{figure}

Experiments are performed on a large-area (mm-scale) MoSe$_2$/WSe$_2$ HS fabricated using a gold tape exfoliation method (see details in Supplementary Note 1.1) \cite{Liu2020}. The HS is prepared with a 4$^{\circ}$, or nearly-aligned, interlayer twist angle, characterized using polarization-resolved second harmonic generation (Supplementary Fig. 1a). We focus on the 4$^{\circ}$ HS in the main manuscript, though experiments were also performed on a nearly anti-aligned MoSe$_2$/WSe$_2$ HS with twist angle of 57$^{\circ}$ (Supplementary Fig. 1b). HS prepared with small twist angles away from either aligned (0$^{\circ}$) or anti-aligned (60$^{\circ}$) have been shown to have strong ILX PL signals \cite{Nayak2017}. A clear signature of the ILX in the 4$^{\circ}$ HS is the low-energy peak in the PL spectrum at $\hbar\omega = 1.35$ eV (Fig. \ref{Fig1}b) to the red of the intralayer PL peaks, which are quenched relative to the PL signals of the individual MLs (Supplementary Fig. 6). Both momentum-direct ($K-K$) and momentum-indirect ($\Sigma-K$) electron-hole transitions are expected to contribute to the PL of the ILX  \cite{Rivera2015,Miller2017c,Hanbicki2018,Rivera2018} which can help to account for the width of the peak in Fig. \ref{Fig1}b. We apply ultrafast TA spectroscopy with a broadband white-light continuum (WLC) probe to study the ILX formation dynamics. We selectively photoexcite the MoSe$_2$ layer by tuning a narrow-band (10 nm) 70-fs pump pulse on resonance with the A exciton peak of MoSe$_2$ ($\hbar\omega~=~1.58~eV$) with a fluence of $3~\mu J/cm^2$ (Supplementary Fig. 2a). For simplicity, when discussing the experimental results, we adopt a notation for the intralayer A and B excitons of each layer using a subscript corresponding to the layer’s transition metal (A$_X$ and B$_X$ where $X$ = Mo or W). The WLC probe beam is generated in a yttrium aluminium garnet (YAG) crystal pumped by a 1 eV pulse and spans a broad spectral window (1.2-2.3 eV) that includes both intralayer excitons and the ILX of the HS (Supplementary Fig. 2a). Unless otherwise stated, the pump and probe pulses are co-circularly polarized such that they access the ($K-K$) optical transitions.

\begin{figure}[h]%
\centering
\includegraphics[width=18cm]{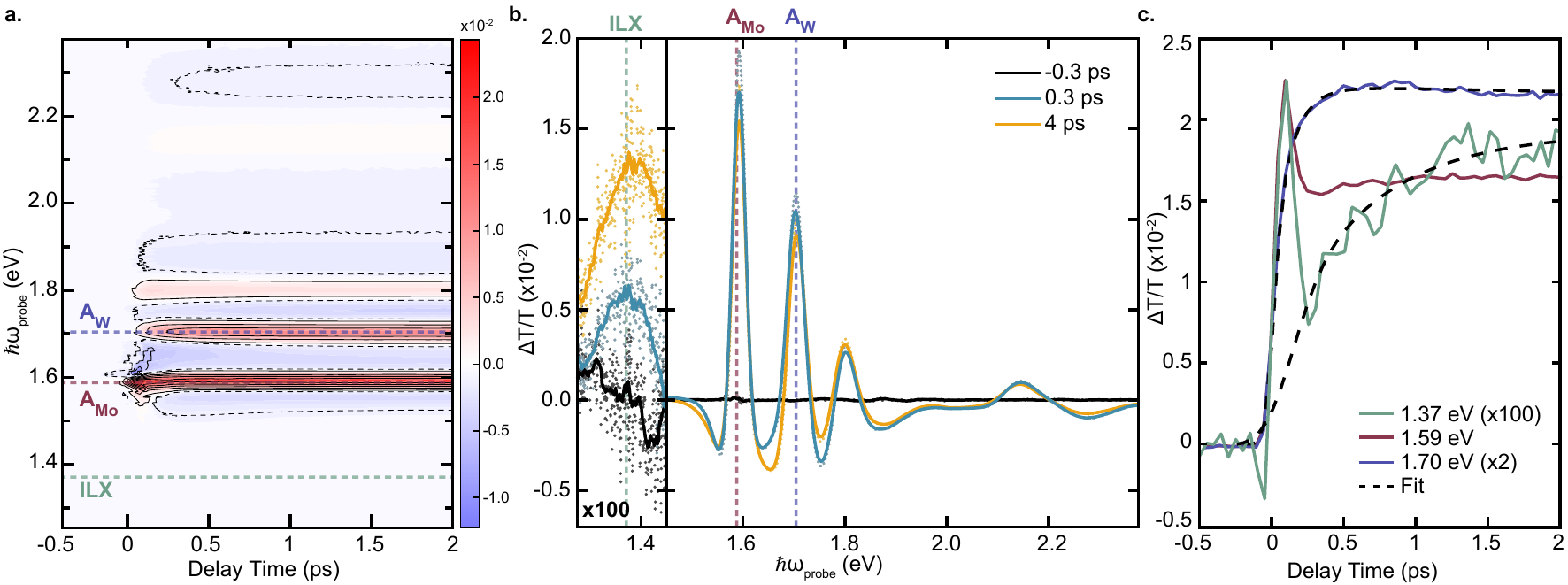}
\caption{\textbf{Transient Optical Response of MoSe$_2$/WSe$_2$ HS.} \textbf{a}, 2D $\Delta T/T$ map as a function of probe energy and delay time. The map displays positive PB (red) and negative PIA signatures (blue) of the intralayer excitons. Dashed lines indicate the probe energy of the three main peaks of interest at A$_W$ (purple), A$_{Mo}$ (red), and the ILX (green). \textbf{b}, $\Delta T/T$ spectra from (\textbf{a}) at selected early time delays. The region below the intralayer exciton optical gap is multiplied by a factor of 100 to highlight the weak ILX peak in the near IR. \textbf{c}, Temporal dynamics of the A$_{Mo}$ (red, $\hbar\omega$ = 1.59 eV), A$_W$ (purple, $\hbar\omega$ = 1.7 eV), and ILX (green, $\hbar\omega$ = 1.37 eV) in the first 2 ps. The A$_W$ and ILX peaks are multiplied by factors of 2 and 100, respectively, to emphasize the delayed rise of these signatures. Dashed lines are the fits to the data.}\label{Fig2}
\end{figure}

Figure \ref{Fig2}a reports a 2D map of the differential transmission ($\Delta T/T$) signal at 77 K as a function of pump-probe delay and probe photon energy. The spectrum is dominated by positive PB signals (red contours) at the energies of the intralayer excitons of both layers (i.e. A$_{Mo}$, A$_W$, B$_{Mo}$, B$_W$) with signal strengths on the order of $\Delta T/T \sim10^{-2}$. The transient signal at the energy of A$_{Mo}$ exciton displays an instantaneous (i.e. pulsewidth-limited) build-up and is the result of the interplay of multiple processes, including quenching of the exciton oscillator strength due to phase-space filling, energy renormalization, and lineshape broadening of the exciton resonance due to Coulomb many-body effects \cite{Trovatello2022}. The same formation dynamics are observed for the B$_{Mo}$ resonance as a consequence of the exchange-driven mixing of the excitons in the ($K-K$) valley \cite{Guo2019b} and light-induced reduction of the Coulomb screening, leading to an instantaneous renormalization of the exciton resonances \cite{Pogna2016}. Following excitation of the A$_{Mo}$ exciton, the A$_W$ resonance shows a delayed PB due to IHT from MoSe$_2$ to WSe$_2$. The finite build-up time of the A$_W$ PB signal (0.20$\pm$0.06 ps, Fig. \ref{Fig2}c) provides a direct estimate of the timescale of the hole scattering process, which is in good agreement with previous observations \cite{Wang2021}. The B$_W$ excitonic resonance shows the same delayed signal formation owing to coupling with the A$_W$ exciton.

Focusing on the ILX energy region of the 2D $\Delta T/T$ map below the optical gap of the intralayer excitons, the TA spectrum at 4 ps pump-probe delay shows a broad peak with a signal strength $\Delta T/T \sim10^{-4}$ (Fig. \ref{Fig2}b). We confirm that this peak is unique to the HS through control measurements on isolated MoSe$_2$ and WSe$_2$ MLs (see Supplementary Note 2.4) and attribute it to PB of the ILX following IHT. The relative strengths of the TA signals of the intralayer excitons and of the ILX are in good agreement with the two orders of magnitude difference in the static transition dipole moments predicted by theory \cite{Yu2015} and measured experimentally \cite{Ross2017c,Barre2022}. Moreover, ILX exhibits valley circular dichroism as reported in Supplementary Note 2.5. We note that the ILX signature is peaked at slightly higher energy in the TA measurements compared to the ILX emission peak in Fig. \ref{Fig1}. We attribute this energy mismatch to the fact that the optically bright ILX signal measured via TA is dominated by momentum-direct ($K-K$) transitions \cite{Barre2022} compared to the ILX PL signal which includes contributions from momentum-direct and momentum-indirect transitions \cite{Okada2018,Hanbicki2018}. Figure \ref{Fig2}c reports the formation dynamics of the ILX. Compared with the dynamics of A$_{Mo}$ and A$_W$ excitons (red and purple, respectively in Fig. \ref{Fig2}c), the ILX TA signal (green) shows an remarkably slower formation timescale (800$\pm$300 fs), significantly longer than the timescale of IHT (200$\pm$60 fs). This behavior suggests that the optically-bright ILX does not form immediately upon IHT and that inferring ILX formation timescales from ICT signatures may be inadequate. The instantaneous peak in the ILX TA signal is due to the weak coherent artifact of the substrate during the pulse overlap, as confirmed by a control measurement on a blank substrate (see Supplementary Note 2.3). This weak coherent signal is not observed at A$_W$ due to its much higher signal strength.

A recent ultrafast TA experiment performed with excitation densities above the exciton Mott transition reported the generation of an interlayer electron-hole plasma whose optical signature consisted of a broad photoinduced absorption (PIA) plateau extending below the optical gap of the HS \cite{Wang2019}. The peak we observe at 1.37 eV is not related to this effect because (i) our excitation density is well below the Mott threshold and (ii) the ILX signal shows a positive signal consistent with a PB signature. At higher excitation densities, we find that the ILX PB signal persists while a negative PIA signal, related to pump-induced modification of the A$_{Mo}$ excitonic resonance \cite{Pogna2016}, increases in strength with fluence until it obscures the weak ILX signature (Supplementary Fig. 8).

\subsection{Theoretical Model}\label{subsec22}

 To understand the observed delayed rise in the ILX signal compared to the IHT process, we developed a microscopic model \cite{Selig2019,Katsch2018} for the exciton dynamics in a perfectly aligned (i.e. 0$^{\circ}$ twist angle) MoSe$_2$/WSe$_2$ HS. The binding energies and the wavefunctions of both intra- and inter- layer excitons have been calculated as the solutions of the Wannier equation for the HS \cite{Holler2021c}. The effective Coulomb potential is determined by solving the Poisson equation for two dielectric slabs in three dielectric environments, following the procedure described in \cite{Ovesen2019h}. Our calculation includes both high energy bound exciton interlayer $s$ states and unbound states above the band edge. A complete picture of the momentum-resolved formation and decay dynamics of excitons is achieved by solving the excitonic Bloch equations for the coherent excitonic polarization and the incoherent excitonic populations \cite{Thranhardt2000}. The pump-induced coherent excitonic polarization, $P^{Mo}$, obtained by excitation slightly above the A$_{Mo}$ resonance, acts as a source term for the incoherent exciton populations via dephasing promoted by exciton-phonon scattering processes. Incoherent population contributions, $N^{Mo/W,\xi_e}$, are referred to with a superscript $\xi$ corresponding to the valley location of the electron where $\xi_e = K/K^{\prime}/\Sigma/\Sigma^{\prime}$ while the hole remains in the $K$ valley; the ILX population is similarly denoted as $N^{IL,\xi_e}$.

\begin{figure}%
\centering
\includegraphics[width=180mm]{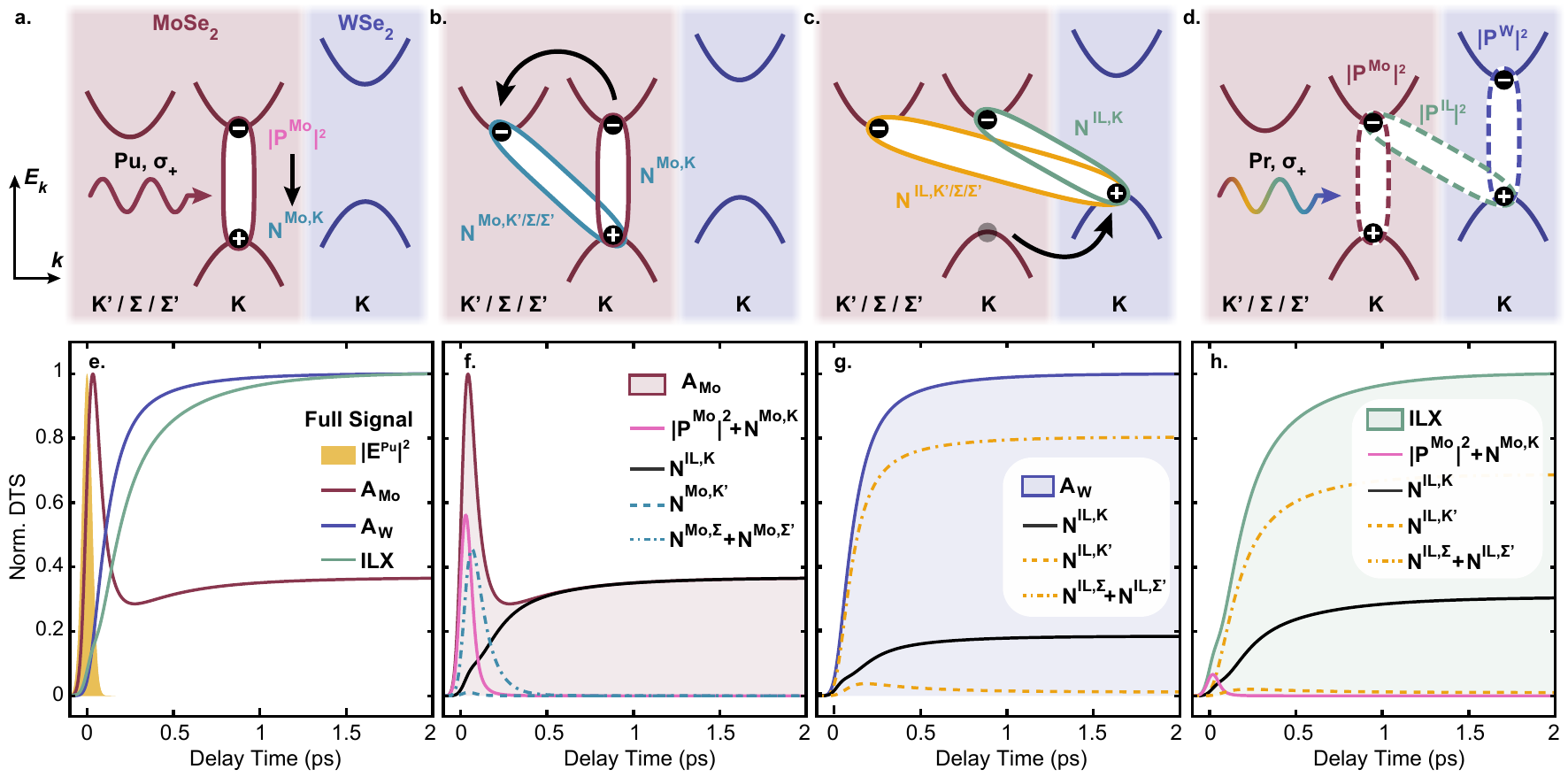}
\caption{\textbf{Sketch of inter- and intra- layer scattering processes and calculated DTS.} \textbf{a}-\textbf{d}, Cartoon diagram of interlayer charge transfer and interlayer exciton formation dynamics. MoSe$_2$ (WSe$_2$) layer and cartoon bands shown in red (purple). The excitons are represented by full ellipses. \textbf{a}, The A$_{Mo}$ exciton is optically pumped initially generating a coherent polarization, $P^{Mo}$, and finally the incoherent $K-K$ exciton, $N^{Mo,K}$. \textbf{b}, Subsequent intervalley electron scattering by phonons leads to the formation of intervalley excitonic populations, $N^{Mo,K^{\prime}/\Sigma/\Sigma^{\prime}}$. \textbf{c}, Phonon-assisted hole tunneling to the WSe$_2$ leads to the formation of momentum direct and indirect ILX populations, $N^{IL,K}$ and $N^{IL,K^{\prime}/\Sigma/\Sigma^{\prime}}$. Note that the $K^{\prime}/\Sigma/\Sigma^{\prime}$ valleys are not equivalent. \textbf{d}, The DTS are calculated for the three probed excitonic transitions (dashed ellipses) highlighted in Fig. \ref{Fig2}. \textbf{e}, Calculated DTS for the A$_{Mo}$, A$_W$ and ILX excitons shown in red, purple, and green, respectively, along with the normalized pump pulse in shaded yellow. All the traces are normalized to their maximum. \textbf{f},\textbf{g},\textbf{h}, Individual intralayer and interlayer excitonic occupations contributing to the DTS in (\textbf{e}).}\label{Fig3}
\end{figure}

In Fig. \ref{Fig3}a-d we report all the scattering processes leading to the formation of the incoherent excitonic populations included in our model. Intraband phonon scattering leads to a sudden decay of exciton polarization, $P^{Mo}$, and nearly instantaneous formation of an intravalley incoherent exciton population at the $K$ valley of MoSe$_2$ layer, $N^{Mo,K}$ (Fig. \ref{Fig3}a). Subsequent intervalley phonon-assisted electron scattering depletes the population, $N^{Mo,K}$, leading to the formation of intervalley excitons within the MoSe$_2$, $N^{Mo,K^{\prime}/\Sigma/\Sigma^{\prime}}$ (Fig. \ref{Fig3}b). To reproduce the experimental results in Fig. \ref{Fig2} it is necessary to include the electron scattering to $\Sigma/\Sigma^{\prime}$ valleys which are located energetically near the $K/K'$ valleys (Supplementary Fig. 4). Finally, interlayer phonon-assisted hole scattering proceeds to the $K$ valley of WSe$_2$, resulting in the formation of momentum direct ILX populations, $N^{IL,K}$, and momentum indirect ILX populations, $N^{IL,K^{\prime}/\Sigma/\Sigma^{\prime}}$ with electrons located at the $K$ or $K^{\prime}/\Sigma/\Sigma^{\prime}$ valleys in the MoSe$_2$ layer, respectively, and holes located at the $K$ valley in the WSe$_2$ layer (Fig. \ref{Fig3}c). 

\begin{figure}%
\centering
\includegraphics[width=88mm]{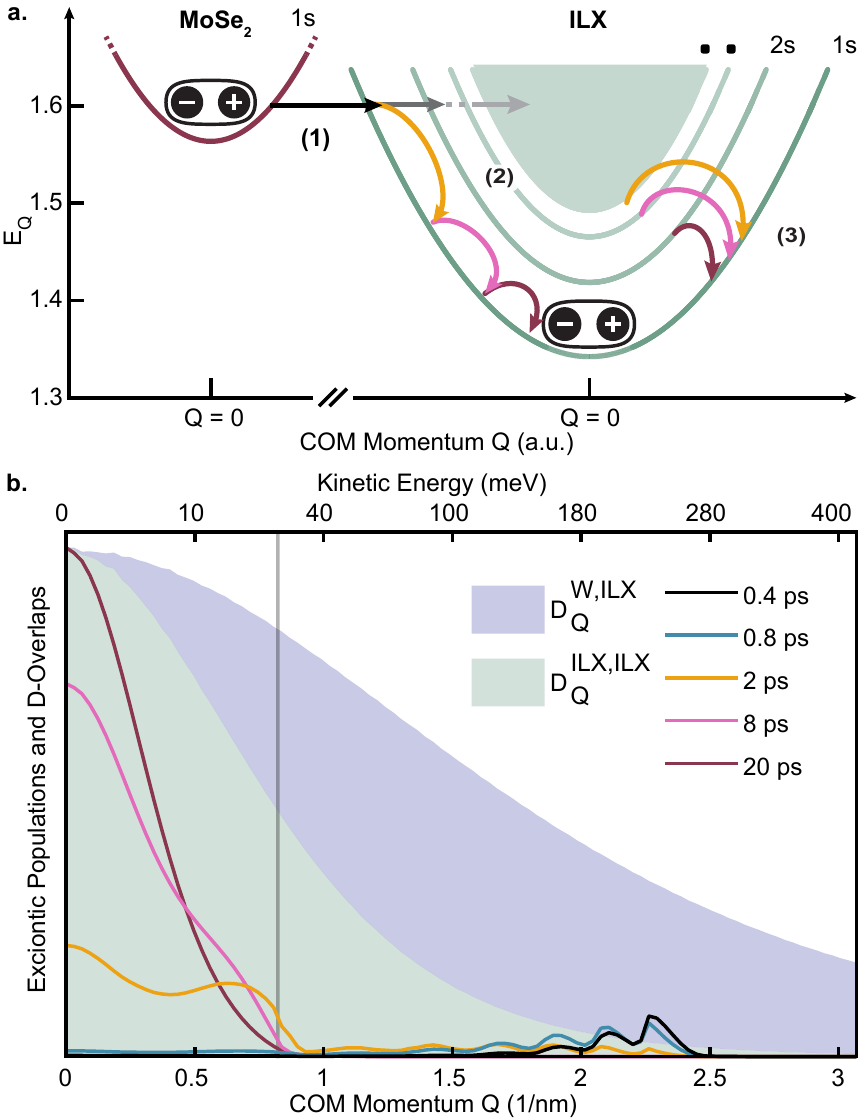}
\caption{\textbf{Hot ILX Relaxation.} \textbf{a}, Sketch of optically bright ILX formation process following thermal relaxation of hot, momentum dark ILX. The formation process occurs in three steps: (1) After resonant excitation of the A$_{Mo}$ transition, the MoSe$_2$ layer is populated by incoherent populations,$N^{Mo,K}_{\textbf Q,1s}$, (left red parabola) and subsequent phonon-assisted IL hole tunneling leads to the formation of ILX incoherent populations, $N^{IL,K}_{\textbf Q,\mu}$, with finite momentum $\textbf Q$ and also high quantum number $\mu$ (right green parabolas) (see Eq.~\eqref{eq:Theory_excitonic_occupations_definition} and Eq.~\eqref{eq:PopulationsSimplifiedNotation}). (2) and (3) The distribution of hot ILX lose energy and momentum by scattering on ultrafast timescales with high energy optical phonons and with low energy acoustic phonons on longer timescales. \textbf{b}, Snapshots of the normalized momentum-dependent excitonic populations of ILX at different delays overlaid on the PB weights of the A$_W$ (solid purple) and ILX (solid green) given by Eq.~\eqref{eq:D_overlaps_e_h}. Following IHT, the population distribution of $K-K$ valley ILX, $N^{IL,K}_{\textbf Q,1s}$, (lines) is peaked around finite Q values between $2-2.5\,\text{nm}^{-1}$, overlapping strongly with A$_W$. The overlap of the $N^{IL,K}_{\textbf Q,1s}$ with the ILX PB weights increases as they shift to lower energy and momentum via scattering processes described in (\textbf{a}). Below the energy threshold of ~$30\,\text{meV}$ (grey line) the scattering is mediated only by acoustic phonons as scattering by optical phonons is energetically forbidden.}\label{Fig4}
\end{figure}

The differential transmission signals (DTS) (Eq.~\eqref{eq:DTS} in the Methods section) are calculated for the probe energies of A$_{Mo}$, (Eq.~\eqref{eq:Theory_DTS_Mprobe}), A$_W$ (Eq.~\eqref{eq:Theory_DTS_Wprobe}), and the momentum-direct ILX (Eq.~\eqref{eq:Theory_DTS_ILprobe}) by performing a separation procedure of the excitonic Bloch equations regarding the probe and pump fields (Fig. \ref{Fig3}d). The DTS temporal traces depend on the interplay of two factors. The first factor is the contribution of coherent polarization and incoherent excitonic populations ($P^{Mo}$, $N^{Mo,\xi_e}$ and $N^{IL,\xi_e}$), where the electron and/or hole part of a given exciton population reduces the absorption of the corresponding optical transition by Pauli blocking and is thus responsible for a positive DTS signal. With this approach, PB signatures by electrons and/or holes are fully taken into account in the excitonic model. The second factor is given by the excitonic PB weights (Eq.~\ref{eq:D_overlaps_e_h} in the Methods section), which emerge as a consequence of the transformation into the excitonic basis and are displayed in Fig.~\ref{Fig4}b for the probed A$_W$ and ILX transitions. Due to these PB weights, the overall excitonic populations do not determine the DTS timetraces directly but rather in a weighted fashion when summed over the center of mass (COM) momenta, \textbf{Q}, and excitonic quantum numbers, $\mu$, (Eq.~\eqref{eq:DTS}). We will examine the PB weights in more detail later on.

The calculated DTS time traces are reported in Fig. \ref{Fig3}e together with the contributions of the individual excitonic populations (Fig. \ref{Fig3}f-h). They qualitatively reproduce the experimental dynamics of intralayer excitons and the ILX, in particular their different formation timescales. The DTS of A$_{Mo}$ exhibits a sharp rise followed by a fast decay and a second delayed rise component (Fig. \ref{Fig3}f). The sharp peak is mainly due to PB by the instantaneous coherent exciton polarization photoexcited at the $K$ valley (pink line, Fig. \ref{Fig3}f) and by the incoherent excitonic populations formed in the MoSe$_2$ layer following phonon-mediated dephasing (blue lines, Fig. \ref{Fig3}). The second delayed rise component originates from the PB contribution of ILX populations (black line, Fig. \ref{Fig3}f) which form following the decay of intralayer exciton populations via phonon-assisted hole tunneling. This bi-exponential behavior of the MoSe$_2$ of the HS is also present in the experimental transient optical response. A comparison of the TA signals, both measured and calculated, for the A$_{Mo}$ exciton of an isolated ML and the HS (see Supplementary Fig. 12) confirms that the transient population of lower-energy ILX state strongly affects the dynamics of the intralayer A$_{Mo}$ state. The DTS of A$_W$ displays a longer formation time than for A$_{Mo}$ as the signal is not influenced by the excitonic intralayer populations in MoSe$_2$ (Fig.~\ref{Fig3}g). Here, only the ILX populations arising from IHT contribute to the PB of the A$_W$ signal.

The DTS of ILX shows similar behavior to the A$_W$ but with a significantly longer risetime (Fig.~\ref{Fig3}h). We attribute this longer formation time to the relaxation dynamics of hot interlayer populations, represented schematically in Fig.~\ref{Fig4}a. Since the binding energies of the ILX ($\approx100\,\text{meV}$) are lower than the difference between the transition energies of A$_{Mo}$ and ILX ($\approx300\,\text{meV}$), phonon-mediated hole transfer creates hot ILX populations, i.e. populations at high momenta $\textbf Q$ and/or higher energy bound and unbound $s$ states of the Rydberg series (Fig.~\ref{Fig4}a), which subsequently scatter to lower-energy momenta and bound states via optical and acoustic phonons. The lower-energy bound states then slowly thermalize into a Boltzmann distribution via scattering with acoustic phonons with energies below the optical phonon bottleneck. Figure \ref{Fig4}b reports the calculated momentum-dependent ILX occupations at different times, taking only the $1s$ state at the $K$ valley into account as an example (the lowest energy parabola in Fig.~\ref{Fig4}a). The ILX occupation is initially peaked at higher momenta following IHT, after which phonon-assisted scattering processes gradually confine the ILX occupations to lower momenta. 

This behavior along with the COM and quantum number dependent PB weights for ILX and A$_W$ transitions (green and purple shaded areas, respectively in Fig.~\ref{Fig4}b) explains the physical origin of the relative delay in the PB signatures of the two transitions. The larger binding energy of the A$_W$ exciton compared to the ILX leads to a broader PB weight distribution, as shown in Fig.~\ref{Fig4}b (see Supplementary Note 1.5). Probing the intralayer A$_W$ transition is more sensitive to the overall excitonic occupation, whereas probing the ILX transition is mainly sensitive to the low-momenta states which are reached at longer delay times. Consequently, only the cold and lowest-energy bound ILX populations significantly contribute to the DTS signal of the ILX transition in Fig. \ref{Fig3}h and the delayed risetime of the ILX compared to the A$_W$ is related to the thermalization of the hot ILX populations. By directly comparing the DTS signals of the probed A$_W$ and ILX transitions, we are able to trace the microscopic mechanisms of the ILX formation and relaxation processes.

\section{Discussion}\label{sec3}

We have used TA spectroscopy with a combination of high sensitivity and broad spectral coverage to directly track the formation dynamics of ILX in a TMD HS. We find that ILX PB signal rises on a picosecond timescale, significantly slower than the build-up dynamics of intralayer exciton PB signals in general and of IHT specifically. Microscopic calculations reproduce the experimental transient signals and explain the formation timescale in terms of different contributions. The hundreds of femtoseconds build-up time of the WSe$_2$ PB signal is mainly related to interlayer scattering of hot holes and therefore represents a rather direct estimation of the hot carrier injection process between the TMD layers of the HS. The slower rise time of the ILX signal is the result of the combination of two scattering processes: (1) phonon-mediated exciton cascade process from unbound and/or higher energy excitonic \textit{s} states to the ground state and (2) intra-exciton energy and momentum relaxation of hot ILX populations. Our simulations also demonstrate that the dynamics of optically bright intra- and inter- layer excitons are influenced by optically dark momentum-indirect excitons.

Besides its fundamental interest, the delayed ILX formation observed and discussed here could explain a long-standing puzzle in optoelectronic devices based on TMD HS, namely the observation of efficient generation of photocurrent despite the large binding energy of the ILX \cite{Bernardi2013,Lee2014,Furchi2014PhotovoltaicHeterojunction,Cheng2014ElectroluminescenceDiodes,Zhang2016}. During their several-hundred femtosecond thermalization process, in fact, the hot ILX have an opportunity to dissociate and form free charge carriers. We foresee that our combined theoretical-experimental approach can be extended to study in real-time exciton formation process in other systems, such as hybrid organic/TMD HS \cite{Zhu2018} and mixed dimensional van der Waals HS \cite{Jariwala2014}.

\section{Methods}\label{sec4}

\subsection{Sample Preparation}\label{subsec41}

Large-area MoSe$_2$/WSe$_2$ HS are prepared using a modified gold tape exfoliation method \cite{Liu2020} resulting in mm-scale ML flakes deposited onto 200 $\mu$m thick transparent SiO$_2$ substrates. TMD HS with varying twist angles are prepared and characterized using polarization-resolved second harmonic generation (Supplementary Fig. 1). We have mainly reported the static and time-domain measurements on a nearly-aligned HS with a 4$^{\circ}$ twist angle. Similar measurements were performed on a nearly-anti-aligned HS characterized by 57$^{\circ}$ twist angle (Supplementary Fig. 14). The HS are prepared with unoverlapped regions where the individual ML can be accessed for control measurements.

\subsection{Transient Absorption Spectroscopy}\label{subsec42}

Time-resolved and static optical measurements are performed at 77 K. TA spectroscopy measurements are performed in a transmission geometry (Supplementary Fig. 2b) with pump and probe beam diameters of 200 $\mu m$ and 100 $\mu m$ at the sample surface, respectively. The relatively large sample region accessed in these TA experiments represents a sort of ensemble measurement where sample heterogeneity contributes to the inhomogeneous broadening, as seen previously on similar samples \cite{Policht2021}. The delay dependent $\Delta T/T$ map and spectra in Fig. \ref{Fig2}a \& b are measured by dispersing the transmitted broadband probe and acquiring it with a fast silicon spectrometer (Entwicklungsbuero EB Stresing) working at 1 kHz laser repetition rate. The probe pulse is a WLC generated by focusing the output of a homemade NIR Optical Parametric Amplifier (OPA) centered at 1 eV into a YAG plate; the WLC spectrum extends from 1.2 to 2.3 eV, covering A/B intralayer and ILX of the HS (Supplementary Fig. 2a). The NIR OPA is seeded with a regeneratively amplified Ti-sapphire laser (Coherent, Libra) emitting 100-fs pulses at 1.55 eV and at a repetition rate of 1 kHz. The pump pulse is generated by a narrowband OPA tuned to the A$_{Mo}$ excitonic resonance ($\hbar\omega$ = 1.59 eV) and modulated at half the repetition rate of the laser. The temporal dynamics of the excitonic resonances (Fig. 2c) are measured by a different TA optical setup based on a Yb:KGW regenerative amplifier (Pharos, Light Conversion) providing 200-fs pulses at 1.2 eV and at a higher repetition rate of 100 kHz. The pump pulse is generated by the second harmonic of a near-IR OPA \cite{Villa2021} tuned to the same energy and fluence conditions as experiments performed on the lower repetition rate TA setup. The WLC is generated by focusing the fundamental of the laser into a YAG crystal. The transmitted probe beam is then sent to a monochromator and a photodiode for lock-in detection. The pump is modulated by a Pockels cell to 50 kHz allowing to reach higher signal-to-noise ratio with respect to the 1 kHz system. The rise times of the TA signals are estimated by fitting the time traces with a rising exponential convoluted with a Gaussian function with full width at half maximum of the instrument response function (FWHM-IRF) of the respective TA setup. The estimated FWHM-IRF is 140 fs for both TA instruments.

\subsection{Theoretical Calculations}\label{subsec44}

To obtain the DTS theoretically, we solve the Bloch equations for the excitonic transitions $P^{i}_{\mu,\textbf Q}=\langle\hat{P}_{\mu,\textbf{Q}}^{i}\rangle $
%$P_{1s}^{\mathrm{Mo/W/ILX}}$
up to third order in the electric field as in \cite{Selig2019}.
Here, $\hat{P}_{\mu,\textbf{Q}}^{i}$ are
the excitonic operators \cite{Katsch2018} carrying excitonic radial quantum
number $\mu$, center of mass momentum $\textbf Q$ and compound layer and valley index $i=\{l_h,\xi_h,l_e,\xi_e\}$
with electron/hole layer $l_{e/h} = M/W$ for MoSe$_2$ and
WSe$_2$, electron valley index $\xi_e = K/K^{\prime}/\Sigma/\Sigma^{\prime}$ and hole valley index $\xi_h =
K$.
Performing a linearization in the weak probe pulse limit \cite{Wegener1990}, we are able to separate the pump induced coherent polarizations
$P^{i}_{\mu,\textbf Q}$
%$P^{\mathrm{Mo}}_{\mu,\textbf Q=\textbf 0}=\langle\hat{P}_{\mu,\textbf{Q}=\textbf 0}^{X=\mathrm{Mo}}\rangle$
	%=\sum_{\textbf q}\varphi_{1s,\textbf q}^{\mathrm{Mo}}v^{\dagger,\mathrm{Mo}}_{\textbf k_h=\textbf q-\beta\textbf Q}c_{\textbf k_e=\textbf q+\alpha\textbf Q}^{\mathrm{Mo}}$
and incoherent populations
\begin{equation}
    N_{\mu,\textbf Q}^{i}=\langle \hat{P}_{\mu,\textbf{Q}}^{\dagger,i} \hat{P}_{\mu,\textbf{Q}}^i\rangle_c
    \label{eq:Theory_excitonic_occupations_definition}
\end{equation}
from the probe-induced dynamics. The coupled equations of motion \cite{Thranhardt2000} are then solved in the time domain. The hole tunneling process is implemented as in \cite{Holler2021c}.
The subscript c in Eq.~\eqref{eq:Theory_excitonic_occupations_definition} accounts
for the purely correlated (or incoherent) part of
the expectation value in the spirit of \cite{Fricke1996,Thranhardt2000}.
By assuming a Dirac delta-shaped probe pulse, we find an expression for the probe- and pump-induced macroscopic polarization $\textbf P^i=\frac{1}{A}\sum_{\mu}\textbf d_{\mu}^iP_{\mu,\textbf Q=\textbf 0}^{i}$ which couples to Maxwell's equations giving the total electric field and therefore the transmission at the HS as in \cite{Hubner1996,Knorr1996}. Here, $\textbf d_{\mu}^i$ is the excitonic dipole moment and $A$ is the illuminated area of the sample. The DTS is then given by subtracting the transmission $T^t$ of the probe pulse without pump pulse from the transmission $T^{t+p}$ of the probe pulse with pump pulse:
$\Delta T(\omega,\tau) = T^{t+p}(\omega,\tau)-T^t(\omega)$ \cite{Malic}. In our case, since we are only interested in timetraces, we neglect Coulomb renormalization and nonlinear broadening \cite{Katsch2020} and focus solely on the Pauli-blocking induced PB dynamics, so that the DTS signal dependent on the time delay $\tau$ between pump and probe pulses can be expressed as
\begin{equation}
    % \begin{split}
        \Delta T^i_{\mu}(\tau)\sim\sum_{i^{\prime},\mathbf Q,\mu}\Big(D_{\mathbf Q,\mu,\nu}^{e,i,i^{\prime}}+D_{\mathbf Q,\mu,\nu}^{h,i,i^{\prime}}\Big)\Big(\big{\vert}P^{i^{\prime}}_{\nu,\mathbf Q}(\tau)\big{\vert}^2\delta_{\mathbf Q,\mathbf 0}+N_{\nu,\mathbf Q}^{i^{\prime}}(\tau)\Big),
    % \end{split}
    \label{eq:DTS}
\end{equation}
where the matrix elements are given by
\begin{equation}
\begin{split}
	D_{\mathbf Q,\mu,\nu}^{e/h,i,i^{\prime}}=&\,\sum_{\mathbf q}\mathbf d_{\mu}^i\cdot\mathbf d_{\mathbf q}^{cv,i}\varphi_{\mu,\mathbf q}^{*,i}\varphi_{\nu,\mathbf q+(-\alpha_{i^{\prime}})/(+\beta_{i^{\prime}})\mathbf Q}^{*,i^{\prime}}\varphi_{\nu,\mathbf q+(-\alpha_{i^{\prime}})/(+\beta_{i^{\prime}})\mathbf Q}^{i^{\prime}}\delta_{\ell_{h/e},\ell_{h/e}^{\prime}}\delta_{\xi_{h/e},\xi_{h/e}^{\prime}}.
 \end{split}
 \label{eq:D_overlaps_e_h}
 \end{equation}
Here, $\textbf d_{\mu}^{i}=\sum_{\textbf q}\varphi_{\mu,\textbf q}^i\textbf d_{\textbf q}^{*,cv,i}$ are the excitonic and $\textbf d_{\textbf q}^{cv,i}$ the electronic dipole moments,
$\varphi_{\mu,\textbf q}^i$ are the excitonic wave functions obtained by solving
the Wannier equation for the respective excitonic
configuration $i$ with radial quantum number $\mu$ and relative
momentum $\textbf q$. $\alpha_i$ and $\beta_i$ are the ratios of the
effective masses \cite{Kormanyos2015a}. In Eq.~\eqref{eq:D_overlaps_e_h}, the indices $\mu$
and $i$ refer to the optically bright probed transition,
whereas $\nu$ and $i^{\prime}$ reflect the bright as well
as dark excitonic populations responsible for the
Pauli blocking. These matrix elements can be viewed as convolutions in momentum space of the excitonic wave functions $\varphi_{\mu,\textbf q}^i$ of the probed transitions $i$ with the excitonic wave functions $\varphi_{\nu,\textbf q}^{i^{\prime}}$ of the pumped populations $i^{\prime}$ and constitute the crucial part of the explanation regarding the different DTS risetimes of the A$_W$ and ILX transition as described in the Results section. The Wannier equation is
solved for a HS using a Coulomb potential
for two dielectric slabs in three dielectric
environments, $\epsilon_1 = 3.9$ (SiO$_2$ substrate on the
MoSe$_2$ side), $\epsilon_g = 1$ (assuming a vacuum environment
between the layers) and $\epsilon_2 = 1$ (vacuum
on the WSe$_2$ side), as in \cite{Ovesen2019h}.

In the following, and to be in line with the notation used in the main part of the manuscript, we denote the excitonic populations under consideration as:
\begin{equation}
    N^{Mo,\xi_e} = N^{M,K,M,\xi_e},~
    %N^{W,\xi_e} = N^{W,K,W,\xi_e},~
    N^{IL,\xi_e} = N^{W,K,M,\xi_e},
    \label{eq:PopulationsSimplifiedNotation}
\end{equation}
and the excitonic coherent polarizations as:
\begin{equation}
    P^{Mo}=P_{1s}^{M,K,M,K},~
    P^{W}=P_{1s}^{W,K,W,K},~
    P^{IL}=P_{1s}^{W,K,M,K}.
\end{equation}
Therefore, the DTS signal for the probed A$_{Mo}$ transition reads explicitely:
\begin{align}
    \begin{split}
        \Delta T^{\text A_{Mo}}(\tau)\sim&\,\sum_{\textbf Q}\Bigg(D_{\textbf Q,1s,1s}^{e+h,Mo,K,Mo,K}\Big(\big{\vert} P^{Mo}(\tau)\big{\vert}^2\delta_{\textbf Q,\textbf 0}+N_{\textbf Q,1s}^{Mo,K}(\tau)\Big)\\
        &\,+\sum_{\xi_e\neq K}D_{\textbf Q,1s,1s}^{h,Mo,K,Mo,\xi_e}N_{\textbf Q,1s}^{Mo,\xi_e}(\tau)+\sum_{\mu}D_{\textbf Q,1s,\mu}^{e,Mo,K,IL,K}N_{\textbf Q,\mu}^{IL,K}(\tau)\Bigg),
    \end{split}
    \label{eq:Theory_DTS_Mprobe}
\end{align}
for the probed A$_W$ transition we obtain:
\begin{align}
    \begin{split}
        \Delta T^{\text A_W}(\tau)\sim\sum_{\textbf Q,\mu,\xi_e}D_{\textbf Q,1s,\mu}^{h,W,K,IL,\xi_e}N_{\mu,\textbf Q}^{IL,\xi_e}(\tau),
    \end{split}
    \label{eq:Theory_DTS_Wprobe}
\end{align}
and for the probed ILX transition it reads:
    \begin{align}
        \begin{split}
            \Delta T^{\text{ILX}}(\tau)\sim&\,\sum_{\textbf Q}\Bigg(D_{\textbf Q,1s,1s}^{e,IL,K,Mo,K}\Big(\big{\vert}P^{Mo}(\tau)\big{\vert}^2\delta_{\textbf Q,\textbf 0}+N_{1s,\textbf Q}^{Mo,K}(\tau)\Big)\\
            &\,+\sum_{\mu}D_{\textbf Q,1s,\mu}^{e+h,IL,K,IL,K}N_{\mu,\textbf Q}^{IL,K}(\tau)+\sum_{\mu,\xi_e\neq K}D_{\textbf Q,1s,\mu}^{h,IL,K,IL,\xi_e}N_{\mu,\textbf Q}^{IL,\xi_e}(\tau)\Bigg).
        \end{split}
        \label{eq:Theory_DTS_ILprobe}
    \end{align}

%Here, we use the following notation regarding layer $l_{e/h}$ and valley index $\xi_{e/h}$ of the corresponding electron-hole pair, where $\mathrm{M/W\xi} = \{l_h=\mathrm{M/W},\xi_h=\mathrm K,l_e=\mathrm{M/W},\xi_e=\xi\}$ and $\mathrm{IL\xi} = \{l_h=\mathrm{W},\xi_h=\mathrm K,l_e=\mathrm{M},\xi_e=\xi\}$. Regarding the notation used in the main part of the manuscript, it holds: $\mathrm{Mo} = \{\mathrm{M},\mathrm K,\mathrm{M},\mathrm K\}$, $\mathrm{W} = \{\mathrm{W},\mathrm K,\mathrm{W},\mathrm K\}$ and $\mathrm{ILX} = \{\mathrm{W},\mathrm K,\mathrm{M},\mathrm K\}$, which correspond to the three probed excitonic transitions.
Eq.~\eqref{eq:Theory_DTS_Mprobe} displays the contribution
of the probed A$_{Mo}$ transition (red
ellipse in Fig.~\ref{Fig3}d) with Pauli blocking due to
electron and hole of the pumped momentum-direct excitonic transitions $P^{Mo}$ and populations $N^{Mo,K}$
(first two terms) as well as due to the blocking
of the hole of momentum-indirect excitonic populations
$N^{Mo,K^{\prime}/\Sigma/\Sigma^{\prime}}$ (third term) and blocking
of the electron of momentum-direct interlayer populations $N^{IL,K}$ (last term).
Eq.~\eqref{eq:Theory_DTS_Wprobe}
describes the probed $A_{W}$ transition
(purple ellipse in Fig.~\ref{Fig3}d) with Pauli blocking
contributions due to holes of all four possible interlayer
populations $N^{IL,\xi_e}$.
Eq.~\eqref{eq:Theory_DTS_ILprobe} shows the probed ILX transition (green ellipse in
Fig. \ref{Fig3}d), where the first two terms account
for Pauli blocking due to the electron of the pumped
momentum-direct excitonic transitions $P^{Mo}$ and populations $N^{Mo,K}$, the third term accounts
for the blocking of electron and hole of
the momentum-direct interlayer population $N^{IL,K}$,
whereas the last term shows the contribution of
the PB by the hole of the other momentum-indirect interlayer
populations $N^{IL,K^{\prime}/\Sigma/\Sigma^{\prime}}$. All Pauli blocking
terms feature distinctive momentum- and excitonic quantum number-dependent
PB weights $D^{e/h,i,i^{\prime}}_{\textbf Q,\mu,\nu}$, which are given in Eq.~\eqref{eq:D_overlaps_e_h}.

\backmatter

\bibliography{mendeley}

\
\bmhead{Supplementary information}
The online version contains supplementary material available including text and figures in sections Extended Methods and Additional Measurements.

\bmhead{Acknowledgments}
We are grateful to C. J. Sayers for helpful discussions and acknowledge F. Morabito, C. Trovatello, A. Genco, S. Sardar and C. D'Andrea for experimental contributions. We would like to thank the IT and data services members of Zuse Institute Berlin for providing the computing infrastructure. S.D.C. acknowledge financial support from MIUR through the PRIN 2017 Programme (Prot. 20172H2SC4). G.C. acknowledge support by the European Union Horizon 2020 Programme under Grant Agreement 881603 Graphene Core 3. X.Y.Z. acknowledges support for sample fabrication by the Materials Science and Engineering Research Center (MRSEC) through NSF grant DMR-2011738. F.S. and A.V. acknowledge support by the European Research Council (ERC) under the European Union’s Horizon 2020 research and innovation programme (grant agreement No. 816313).
We acknowledge financial support from the Deutsche Forschungsgemeinschaft (DFG) through SFB 951 Project No. 182087777 (M.K., M.S. and A.K) and Project KN 427/11-1 (H.M. and A.K.) Project No. 420760124.

\section*{Additional Information}

\begin{itemize}
\item Competing interests: 
The authors declare no competing interests.
%\item Ethics approval 
%Not applicable
%Not applicable
%\item Consent for publication
\item Availability of data and materials:
Data pertaining to this work will be made available by request to the corresponding authors.
%Not applicable
\item Authors' contributions:
V.R.P and S.D.C. devised the experimental work. V.R.P., O.D., and A.V. performed the experimental work; V.R.P. and O.D. performed data analysis. H.M., M.K., and B.K. performed the theoretical work. Q.L. prepared and characterized the TMD HS samples. All authors contributed to the interpretation of the results and preparation of the manuscript.
\end{itemize}

\noindent
% If any of the sections are not relevant to your manuscript, please include the heading and write `Not applicable' for that section. 

% \bibliography{mendeley}% common bib file

\newpage

\clearpage

\newpage

\clearpage 

\end{document}